\documentclass[11pt,a4paper]{article}
\usepackage{graphicx}% Include figure files
\usepackage{dcolumn}% Align table columns on decimal point
\usepackage{bm}% bold math
\def\nn {\nonumber}

\newcommand{\be}{\begin{equation}}
\newcommand{\ee}{\end{equation}}
\newcommand{\bea}{\begin{eqnarray}}
\newcommand{\eea}{\end{eqnarray}}

\newcommand{\ep}{\epsilon}

\newcommand{\om}{\omega}

\newcommand{\ov}{\overline}

\newcommand{\qs}{q \!\!\! /}
\newcommand{\ks}{k \!\!\! /}

\newcommand{\tom}{\widetilde{\om}}   

\newcommand{\vk}{\vec k}

\newcommand{\vq}{\vec q} 
\newcommand{\vl}{\vec l}

\newcommand{\mn}{\mu\nu}
\newcommand{\del}{\partial}

\begin{document}

\title{A real-time thermal field theoretical analysis of Kubo-type Shear viscosity :
Numerical understanding with simple examples}

\author{Sabyasachi Ghosh}
\date{}
\maketitle
\begin{center}
{$^1$Instituto de Fisica Teorica, Universidade Estadual Paulista, 
Rua Dr. Bento Teobaldo Ferraz, 271, 01140-070 Sao Paulo, SP, Brazil}
\end{center}

%\begin{history}
%\received{Day Month Year}
%\revised{Day Month Year}
%\end{history}

\begin{abstract}
A real-time thermal field theoretical calculation of shear viscosity has been described 
in the Kubo formalism for bosonic and fermionic medium.
The two point function of viscous stress tensor in the lowest order 
provides one-loop skeleton diagram of boson or fermion field for bosonic
or fermionic matter. According to the traditional diagrammatic technique of transport
coefficients, the finite thermal width of boson or fermion is
introduced in their internal lines during the evaluation of boson-boson
or fermion-fermion loop diagram. These thermal widths of $\phi$ boson 
and $\psi$ fermion are respectively obtained from the imaginary part of self-energy
for $\phi\Phi$ and $\psi\Phi$ loops, where interactions of higher mass
$\Phi$ boson with $\phi$ and $\psi$ are governed by the simple $\phi\phi\Phi$ and
${\ov\psi}\psi\Phi$ interaction Lagrangian densities.
A two-loop diagram, having same power of coupling constant as in the one-loop
diagram, is deduced and its contribution appears much lower than the one-loop values
of shear viscosity. Therefore the one-loop results of Kubo-type shear viscosity
may be considered as leading order results for this simple $\phi\phi\Phi$ and
${\ov\psi}\psi\Phi$ interactions. This approximation is valid
for any values of coupling constant and at the 
temperatures greater than the mass of constituent particles of the medium.
\end{abstract}

%\pacs{12.38.Mh,25.75.-q,24.85.+p,25.75.Nq}
\maketitle
\section{Introduction}
A medium can be perturbed slightly away
from its equilibrium state due to quantum fluctuations 
at finite temperature. The responses to these
fluctuations can be determined from the dissipative
quantities of the medium.
The different fluxes originating from their corresponding thermodynamical forces 
can appear in the medium to restore the system to its equilibrium 
state. The transport coefficients like shear and bulk viscosities are proportional
constants of the linear relations between these fluxes and forces.

With the help of Kubo formula, the transport coefficients can be
microscopically estimated from their corresponding retarded thermal correlators in
the low frequency limit.
For example, the shear viscosity coefficient can be obtained from the 
retarded correlator of the viscous stress tensor. The simplest possible
anatomical representation of the correlator is a one-loop kind of
diagram, where
a finite thermal width has to be included in the propagator
of the constituent particle.
This is a very well established 
technique~\cite{Hosoya,Horsley,Nicola,Nicola2,Nicola3,Weise,Weise2},
which is generally adopted
to get a non-divergent value of the transport coefficient.
Beyond the one-loop, there will be infinite class of planar ladder diagrams, 
which have the same power of coupling constant
as in the one-loop diagram. The reason is that the
additional coupling constants from the interaction vertices of any higher 
order diagram are exactly cancelled by the additional thermal widths
coming from the each loop of that diagram. For this purpose, different ways
of re-summed technique are proposed for $\phi^4$ 
theory~\cite{Jeon,Wang,Carrington,Carrington2},
as well as for hot QCD~\cite{Basagoiti}. In this context,
a Kubo-type shear viscosity calculation for medium with
$\phi$ boson or $\psi$ fermion has been derived
in a generalized manner where a simple
$\phi\phi\Phi$ or $\psi\psi\Phi$ interaction Lagrangian is
considered. The $\Phi$ boson is assumed as a higher mass intermediate particle,
which is supposed to be appeared in the elastic $\phi\phi$ or $\psi\psi$ scattering.
We can take $\pi\pi\sigma$~\cite{GKS} or $QQ\pi$~\cite{Ghosh_NJL} 
interaction as a practical example of such a case.

At first, the simplest possible diagrammatic representation of
shear viscosity as a $\phi\phi$ or $\psi\psi$ loop
is derived explicitly in the real-time thermal field theory (RTF),
which is addressed in Sec.~(2). For getting a non-divergent
shear viscosity, a finite thermal width of $\phi$ or $\psi$
has to be introduced owing to the traditional diagrammatic calculation
of transport coefficient. In Sec.~(3), the thermal width of $\phi$
or $\psi$ is deduced from the imaginary part of its thermal self-energy
for $\phi\Phi$ or $\psi\Phi$ loop. 
The possibilities of different infinite number of diagrams 
for the $\phi\phi\Phi$ interaction are pointed out in Sec.~(4),
where a two-loop diagram is explicitly derived in RTF.
During numerical estimations of one-loop
and two-loop contributions in shear viscosity, two-loop is substantially
suppressed due to presence of additional number of thermal distribution
functions appearing in the multiplicative way. These numerical results are
discussed in Sec.~(5), and Sec.~(6) concludes the article. 
\section{Formalism}
From the simple derivation of Kubo formula~\cite{Zubarev,Kubo},
we are starting our calculation with the expression of shear viscosity in momentum
space~\cite{Nicola},
\be
\eta=\frac{1}{20}\lim_{q_0,\vq \rightarrow 0}\frac{A_\eta(q_0,\vq)}{q_0}~,
\label{eta_Nicola}
\ee
where 
\be
A_\eta(q_0,\vq)=\int d^4x e^{iq\cdot x}\langle[\pi_{ij}(x),\pi^{ij}(0)]\rangle_\beta
\ee
is spectral function of space component of the viscous-stress tensor, 
$\pi^{\mn}$ and $\langle \hat{O}\rangle_\beta$ for any operator $\hat{O}$
denotes the equilibrium ensemble average,
\bea
\langle \hat{O}\rangle_\beta&=&{\rm Tr}\frac{e^{-\beta H}\hat{O}}{Z}
\nn\\
{\rm with}~~ Z&=&{\rm Tr}e^{-\beta H}~.
\eea
The $\pi^{\mn}$ comes from the dissipative
part of total energy-momentum tensor, $T^{\mn}$. In the relativistic dissipative
hydrodynamics~\cite{Romatschke_rev,Hirano} the $T^{\mn}$ 
can be decomposed into an ideal part $T_0^{\mn}$ and a 
dissipative part $T_D^{\mn}$. 
\be
{\rm\it i.e.}~~ T^{\mn}=T_0^{\mn}+T_D^{\mn}~,
\ee
where
\bea 
T_0^{\mn}&=&\epsilon u^\mu u^\nu - P\Delta^{\mn}
\nn\\
{\rm and}~~ T_D^{\mn}&=& K^\mu u^\nu + K^\nu u^\mu- \pi^{\mn}~,
\label{Tmn_pimn}
\eea
with $\Delta^{\mn}=g^{\mn}-u^\mu u^\nu$ and $\epsilon, P, u^\mu$ are respectively 
energy density, pressure density, four velocity of the matter. 
By applying appropriate projection operators,
the candidates of right hand side of the Eq.~(\ref{Tmn_pimn}) can be
extracted from $T^{\mn}$ as 
\be
\left\{ \begin{array}{llll}
\epsilon
\\
P
\\
K^\rho
\\
\pi^{\rho\sigma}
\end{array}
\right\}
=\left\{ \begin{array}{llll}
u^\mu u^\nu 
\\
- \frac{1}{3}\Delta^{\mn}
\\
\Delta^{\rho\mu}u_{\nu}
\\
(\Delta^{\rho\sigma}\Delta^{\mn}-\frac{1}{3}\Delta^{\rho\mu}\Delta^{\sigma\nu})
\end{array}
\right\}
T_{\mn}.
\ee
For $\phi$ boson and $\psi$ fermion the viscous-stress tensor 
can be deduced as (see Appendix A)
\be
\pi_{\mn}=(\Delta^\rho_\mu\Delta^\sigma_\nu-\frac{1}{3}\Delta_{\mn}\Delta^{\rho\sigma})
\left\{\begin{array}{ll}
\del_\rho\phi\del_\sigma\phi
\\
i{\ov \psi}\gamma_\rho\del_\sigma\psi
\end{array}
\right\}~.
\label{pimn}
\ee
\begin{figure}
\begin{center}
\includegraphics[scale=0.5]{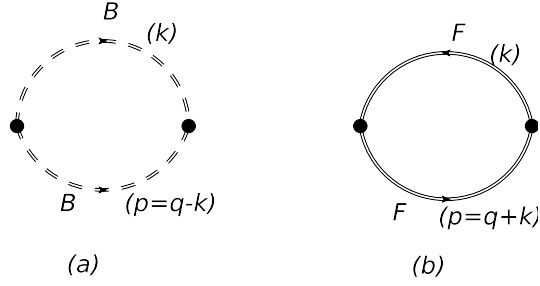}
\caption{Boson-Boson (a) and Fermion-Fermion (b) loop diagram 
after Wick contraction from two point function of corresponding viscus-stress tensor.}
\label{diagram}
\end{center}
\end{figure}
In RTF, the thermal average of any two point function
always give a $2\times 2$ matrix structure. Hence the two point correlator
of viscous-stress tensor will be
\be
\Pi_{ab}(q)=i\int d^4x e^{iqx}\langle T_c\pi_{\mn}(x)\pi^{\mn}(0)\rangle^{ab}_\beta~,
\label{pi_ab}
\ee
where the superscripts $a, b (a,b=1,2)$ represent the (thermal) indices
of the matrix and $T_c$ denotes time ordering with respect to a contour
in the complex time plane. Of the variety of possible contours in the 
complex time plane~\cite{Umez}, two are specially interesting,
namely the closed one~\cite{Keldysh} and the symmetrical one
~\cite{Semenoff, SS_SM}. Here the latter contour
is chosen .

The matrix can be diagonalized in terms of 
a single analytic function which determines completely the dynamics of the 
corresponding two point function. The diagonal element can also be related
with the retarded two point function of viscous-stress tensor.
These retarded function, $\Pi^R(q)$ (say) or diagonal element, ${\ov \Pi}(q)$
(say) is simply related to any one component, say 11 component of the matrix.
The relation among the imaginary part of these components 
(for $\phi\phi$ or $\psi\psi$ loops) is
\be
{\rm Im}\Pi^R(q)=\epsilon(q_0){\rm Im}{\ov \Pi}(q)
={\rm tanh}(\frac{\beta q_0}{2})
{\rm Im}\Pi_{11}(q)~.
\label{R_bar_11}
\ee
%where $\mu^B_q$ and $\mu^F_q$ are respectively chemical
%potential of B and F.
Now the imaginary part of retarded two point function of any
field quantity is directly related with the corresponding
spectral function. For our present  case, the viscous-stress tensor 
$\pi_{\mn}$ is the field quantity. Hence the relation between
corresponding spectral function $A_\eta(q)$ and retarded function
$\Pi^R(q)$ is given by
\be
A_\eta(q)=2{\rm Im}\Pi^R(q)~.
\ee
Using this relation, the Eq.~(\ref{eta_Nicola})) become
\be
\eta=\frac{1}{10}\lim_{q_0,\vq \rightarrow 0}
\frac{{\rm Im}\Pi^R(q_0,\vq)}{q_0}~.
\label{eta_Im_R}
\ee
To calculate this retarded two point function, let us start from
11 component of the $\Pi_{ab}$ matrix. After using Eq.~(\ref{pimn}) in 
Eq.~(\ref{pi_ab}), the Wick contraction of the
boson fields ($\phi$) and fermion fields ($\psi$) generate 
an one-loop kind of skeleton diagram, which are shown 
in Fig.~\ref{diagram}(a) and (b) respectively. 
They can be mathematically expressed
as~\cite{S_rev}
\be 
\Pi_{11}(q)=i\ep_F\int \frac{d^4k}{(2\pi)^4}N(q,k) D_{11}(k)
D_{11}(p)~,
\label{self_eta}
\ee 
where $D_{11}(k)$ and $D_{11}(p)$ are the scalar part of the
propagators at finite temperature. Now $p=q-k$ for boson-boson
($\phi\phi$) loop whereas $p=q+k$ for fermion-fermion ($\psi\psi$) loop.
In fermion loop there must be an extra negative sign according
to Feynman rule, which is maintained by putting 
$\ep_F=+,-$ for $\phi\phi$ and $\psi\psi$ loops respectively.
The term $N(q,k)$ contain
the numerator part of two propagators as well as the vertex
kind of part which is determined from Eq.~(\ref{pimn}). 
The expression of $N(q,k)$ (multiplying with the $\epsilon_F$) 
for $\phi\phi$ and $\psi\psi$ loop are respectively given by (see Appendix B)
\be
\epsilon_FN(q,k)=-[\vk^2(\vq-\vk)^2+\frac{\{\vk\cdot(\vq-\vk)\}^2}{3}]
\label{N_qvkv_BB}
\ee
and
\be
\epsilon_FN(q,k)=-\frac{32}{3}\{k_0(q_0+k_0)\}\{\vk\cdot(\vq+\vk)\}
+4[\{\vk\cdot(\vq+\vk)\}^2+\frac{\vk^2(\vq+\vk)^2}{3}]~.
\label{N_qvkv_FF}
\ee
Just by absorbing the $\ep_F$ in $N(q,k)$ we will not write it
for our further calculations.

In RTF, the 11 component
of the scalar part of the thermal propagator
can be expressed as
\bea
D^{11}(k)&=&\frac{-1}{k_0^2-\om_k^2+i\ep}+2\pi i\ep_k
F_k(k_0)\delta(k_0^2-\om_k^2),
~{\rm with}~ F_k(k_0)=n^+_k\theta(k_0)+
n^-_k\theta(-k_0)
\nn\\
&=&-\frac{1}{2\om_k}(\frac{1-n^+_k}{k_0-\om_k+i\ep}+
\frac{n^+_k}{k_0-\om_k-i\ep}
-\frac{1-n^-_k}{k_0+\om_k-i\ep}
-\frac{n^-_k}{k_0+\om_k+i\ep})~.
\label{de11}
\eea
Here $n^{\pm}_k(\om_k)=\frac{1}{e^{\beta(\om_k \mp \mu)}-\ep_k}$
($\ep_k=+1,-1$ for $\phi$ and $\psi$ respectively)
is the thermal distribution of $\phi$ with energy $\om_k=(\vk^2+m_\phi^2)^{1/2}$
or $\psi$ with energy $\om_k=(\vk^2+m_\psi^2)^{1/2}$,
where the $\pm$ sign in the superscript of $n_k$ refer to particle and 
anti-particle respectively.  
%Here $\mu$ is the chemical potential associated with Fermi energy
%for F propagator whereas for B propagator it will be zero. 
%In the second line of Eq.~(\ref{de11}) for F propagator, 
%the first and the second terms 
%are associated with the propagation
%above the Fermi sea and holes in the Fermi sea respectively while
%the third and
%fourth terms represent the corresponding situations for
%anti-particles. 
%The full relativistic quark propagator thus treats the
%particle and anti-particle on an equal footing and the all possible singularities 
%(particle, hole of the particle, anti-particle, hole of the anti-particle) 
%are automatically included. 

After doing the $k_0$ integration of Eq.~(\ref{self_eta}) and then using the
relations (\ref{R_bar_11}), we get the imaginary part of 
retarded self-energy~\cite{Ghosh_thesis}
\bea
{\rm Im}{\Pi}^{R}(q)&=&\int\frac{d^3k}{(2\pi)^3}
\frac{(-\pi)N}{4\om_k\om_p}[C_1\delta(q_0 -\om_k-\om_p)
+C_2\delta(q_0-\om_k+\om_p)
\nn\\
&&+C_3\delta(q_0 +\om_k-\om_p)
+C_4\delta(q_0 +\om_k+\om_p)]~,
\label{Pi_LU}
\eea
where $\om_p=\sqrt{(\vq\mp\vk)^2+m^2}$ ($\mp$ for $\phi\phi$ and $\psi\psi$
loop respectively) and $N=N(k_0=\pm\om_k,\vk,q)$ .
Here $C_i (i=1,..,4)$ are four different statistical probability 
attached with four different delta functions.
For $\phi\phi$ loop, $C_i (i=1,..,4)=\{1+n^+_k(\om_k)+n^+_p(q_0-\om_k)\}$, 
$\{-n^+_k(\om_k)+n^-_p(-q_0+\om_k)\}$, $\{n^-_k(\om_k)-n^+_p(q_0+\om_k)\}$, 
$\{-1-n^-_k(\om_k)-n^-_p(-q_0-\om_k)\}$
whereas $C_i (i=1,..,4)=\{-1+n^-_k(\om_k)+n^+_p(q_0+\om_k)\}$, 
$\{-n^-_k(\om_k)+n^-_p(-q_0+\om_k)\}$, 
$\{n^+_k(\om_k)-n^+_p(q_0+\om_k)\}$, 
$\{1-n^+_k(\om_k)-n^-_p(-q_0-\om_k)\}$
for $\psi\psi$ loop.
%From Eq.~() $N(k_0)$ is independent of $k_0$, so
%all $N_i$'s are same, say $N$.
%The different region of the discontinuities of the complex function 
%${\Pi}^{R}(q)$ in the real axis of complex $q_0$ plane will be associated with
%the non-zero magnitude of the ${\rm Im}{\Pi}^{R}(q)$. 
Four different branch cuts in the real axis of 
complex $q_0$ plane can be identified
from four different terms of ${\rm Im}{\Pi}^{R}(q)$ with different delta functions.
In addition to the unitary cut,
present already in vacuum, the thermal field theory calculation generates a new cut, 
the so called the Landau cut. The region of these different branch cuts are
($-\infty$ to $-\sqrt{\vq^2+4m_{\phi,\psi}^2}$) for unitary cut in negative $q_0$-axis,
($-\vq$ to $\vq$) for Landau cut and 
($\sqrt{\vq^2+4m_{\phi,\psi}^2}$ to $\infty$) for unitary cut in positive $q_0$-axis.
%=============   ============================================
%Here $\Gamma$ for free propagator is a very small quantity and
%tending to zero. It has a role just for infinitesimal shifting 
%the pole position of the propagator in complex $k_0$ plane.
%For exact propagator $\Gamma$ is not limiting quantity, tending
%to zero. A finite small value of $\Gamma$ can be introduced through
%Dyson's series expansion of a propagator.
%========================   ===================================
Now to calculate $\eta$ from the ${\Pi}^{R}(q)$, we have to
concentrate on the limit of $q_0, \vq\rightarrow 0$. Therefore the Landau
cuts will be only the focus of our calculation.
Hence using the Landau part of Eq.~(\ref{Pi_LU}) in Eq.~(\ref{eta_Im_R}), we have
\bea
\eta&=&\frac{1}{10}\lim_{q_0,\vq \rightarrow 0}\frac{1}{q_0}
\int\frac{d^3k}{(2\pi)^3}\frac{(-\pi)N}{4\om_k\om_p}
\{C_2\delta(q_0-\om_k+\om_p)
+C_3\delta(q_0+\om_k-\om_p)\}
%\right.\nn\\
%&&~~~~+\left.\frac{-n^-_k  +n^-_p }{q_0 -\om_k+\om_p-i\Gamma}\}\right]
\nn\\
&=&\frac{1}{10}\lim_{q_0,\vq \rightarrow 0}{\rm Im}
\left[\int\frac{d^3k}{(2\pi)^3}\frac{N}{4\om_k\om_p}\lim_{\Gamma \rightarrow 0}
\right.\nn\\
&&\left.
\left\{\frac{C_2/q_0}{(q_0-\om_k+\om_p)+i\Gamma}
+\frac{C_3/q_0}{(q_0+\om_k-\om_p)+i\Gamma}\right\}\right]~.
\label{eta_Gama}
\eea
To get a non-divergent
contribution of $\eta$, the further calculation will be continued 
for finite value of $\Gamma$. 
This is the place where the interaction scenario is introduced, which is
very essential for a dissipative system. This is done
by transforming the delta functions for free theory to the spectral
functions with finite thermal width for interacting theory. 
The $\Gamma$ is identified as the thermal width (or collision rate)
of the constituent particles and it reciprocally measures the dissipative
coefficients like the shear viscosity. In this respect this formalism
may be equivalent to quasi particle approximation and this trained is widely
used for calculating the
dissipative coefficients in Kubo approach~\cite{Nicola, Weise}.
Using the limiting values,
\be
\lim_{\vq \rightarrow 0}\om_p=\om_k
\ee
in Eq.~(\ref{eta_Gama}), we have
\be
\eta=\frac{1}{10}\int\frac{d^3k}{(2\pi)^3}\frac{(-N^0)}{4\om_k^2\Gamma}[
I_2+I_3]~,
\label{eta_I}
\ee
where 
\be
N^0=\lim_{q_0, \vq\rightarrow 0}N(k_0=\pm\om_k,\vk,q)
\label{N_0}
\ee
and
\be
I_{2,3}=\lim_{q_0 \rightarrow 0}\frac{C_{2,3}(q_0)}{q_0}~,
\label{I_23}
\ee
with
\be
C_{2,3}(q_0)=\mp n^{\pm}_k(\om_k)\pm n^{\mp}_p(\mp q_0+\om_k)~.
\ee
Here in Eq.~(\ref{I_23}), we see that the limiting value
of $I_{2,3}$ is appeared to be a $0/0$ form and
so we can apply the L'Hospital's rule i.e.
\bea
I_{2,3}&=&\lim_{q_0\rightarrow 0}\frac{\frac{d}{dq_0}
\{C_{2,3}(q_0)\}}{\frac{d}{dq_0}\{q_0\}}
\nn\\
&=&\beta n_k^{\mp}(1+\ep_kn^{\mp}_k)~,
\label{I}
\eea
since
\bea
\frac{d}{dq_0}\{\pm n^{\mp}_p(\om_q=\mp q_0+\om_k)\}&=&\pm\frac{-\beta \frac{d\om_q}{dq_0} 
e^{\beta(\om_q\pm\mu)}}{\{e^{\beta(\om_q\pm\mu)}-\ep_k\}^2}
\nn\\
\lim_{q_0\rightarrow 0}\frac{d}{dq_0}\{\pm n^{\mp}_p(\om_q=\mp q_0+\om_k)\}
&=&\pm\frac{-\beta (\mp) e^{\beta(\om_k\pm\mu)}}{\{e^{\beta(\om_k\pm\mu)}-\ep_k\}^2}
\nn\\
&=&\beta [n^{\pm}_k(1+\ep_kn^{\pm}_k)]~.
\eea
Depending on the sign of $\ep_k$, the statistical probability
become Bose enhanced ($\ep_k=+1$ for $\phi$) or Pauli blocked
($\ep_k=-1$ for $\psi$) probability.
Hence using Eq.~(\ref{I}) in Eq.~(\ref{eta_I}), we can get a generalized expression
of shear viscosity coefficient for $\phi$ boson or $\psi$ fermion,
\bea
\eta&=&\frac{\beta}{10}\int\frac{d^3k}{(2\pi)^3}\frac{(-N^0)}{4\om_k^2\Gamma}
[n^-_k(1+\ep_kn^-_k)+n^+_k(1+\ep_kn^+_k)]
\nn\\
&=&\frac{\beta}{80\pi^2}\int \frac{\vk^2d\vk}{\om_k^2}
\frac{(-N^0)}{\Gamma}[n^-_k(1+\ep_kn^-_k)
+n^+_k(1+\ep_kn^+_k)]~.
\label{eta_last}
\eea
From the final expression we can understand the importance of finite value
of $\Gamma$ to get a non-divergent value of $\eta$.  
A finite small value of $\Gamma$ can be introduced through
Dyson's series expansion of a propagator and we can repeat our calculation from
Eq.~(\ref{de11}) by using $i\Gamma$ in place of $i\eta$.
The corresponding $\Gamma$ for $\phi$ and $\psi$ can be quantitatively defined as
\be
\Gamma_\phi=-{\rm Im}{\Pi}^R_{(\phi)}(k_0=\om_k,\vk)/m_\phi~~~~~({\rm for~}\phi)~,
\label{Gam_B}
\ee
\be
\Gamma_\psi=-{\rm Im}\Sigma^R_{(\psi)}(k_0=\om_k,\vk)~~~~~~~~~~({\rm for~}\psi)~,
\label{Gam_F}
\ee
where ${\rm Im}{\Pi}^R_{(\phi)}(k)$ and ${\rm Im}\Sigma^R_{(\psi)}(k)$ are the imaginary
part of retarded self-energy for $\phi$ and $\psi$ 
respectively, which will be even functions of $k_0$.

Let us now concentrate on the Eq.~(\ref{N_0}) to deduce $N^0$. 
In this limiting case ($q_0,\vq\rightarrow 0$), the
Eq.~(\ref{N_qvkv_BB}) and (\ref{N_qvkv_FF}) can be simplified as
\be
N^0=-\frac{4\vk^4}{3}~~~ {\rm for~}\phi\phi~{\rm loop}
\ee
and
\be
N^0=-\frac{16\vk^4}{3}~~~ {\rm for~}\psi\psi~{\rm loop}~.
\ee
Hence the shear viscosity of the medium with bosons ($\phi$)
and fermions ($\psi$) are respectively given below
\bea
\eta_\phi&=&\frac{\beta }{30\pi^2}\int^{\infty}_{0} 
\frac{d\vk\vk^6}{\om_k^2\Gamma_\phi}\frac{\{n^+_k(1+n^+_k)+n^-_k(1+n^-_k)\}}{2}
~~~{\rm for~}\phi\phi~{\rm loop}
\nn\\
&=&\frac{\beta }{30\pi^2}\int^{\infty}_{0} 
\frac{d\vk\vk^6}{\om_k^2\Gamma_\phi}n_k(1+n_k) 
~~~~~~~({\rm at}~\mu=0 )
\label{eta_BB}
\eea
and
\be
\eta_\psi=\frac{\beta }{15\pi^2}\int^{\infty}_{0} 
\frac{d\vk\vk^6}{\om_k^2\Gamma_\psi}\{n^+_k(1-n^+_k)+n^-_k(1-n^-_k)\}
~~~{\rm for~}\psi\psi~{\rm loop}~.
\label{eta_FF}
\ee
{\bf Some special cases :} Now taking isospin degeneracy factor $I_\pi=3$ for pionic medium 
we can generate the
expression of ~\cite{Nicola,Weise,SS_SM,S_rev},
\be
\eta_\pi=\frac{\beta}{10\pi^2}\int^{\infty}_{0} 
\frac{d\vk\vk^6}{\om_k^2\Gamma_\pi}n_k(1+n_k)~,
\ee
where $n_k=n^{\pm}_k$ for $\mu_\pi=0$ has to be considered.
This one-loop expression of shear viscosity
from Kubo approach~\cite{Nicola,Weise,S_rev} is exactly coincide with the
expressions from the relaxation-time approximation of
the kinetic theory approach~\cite{Gavin,Prakash,SSS,S_rev}.

When the constituent particles are nucleon, then we have to take
isospin degeneracy $I_N=2$ in Eq.~(\ref{eta_FF}) i.e.
\be
\eta_N=\frac{2\beta }{15\pi^2}\int^{\infty}_{0} 
\frac{d\vk\vk^6}{\om_k^2\Gamma_N}\{n^+_k(1-n^+_k)+n^-_k(1-n^-_k)\}~.
\ee
Similarly for 2-flavor ($N_f=2$) quark matter the degeneracy factor become
$N_fN_c=6$ (as color degeneracy $N_c=3$) and so
\be
\eta_Q=\frac{2\beta }{5\pi^2}\int^{\infty}_{0} 
\frac{d\vk\vk^6}{\om_k^2\Gamma_Q}\{n^+_k(1-n^+_k)+n^-_k(1-n^-_k)\}~.
\ee
This expression is also exactly coincide with the
expressions from the relaxation-time approximation~\cite{Redlich}.

\section{Calculation of $\Gamma$}
Let us take a simple $\phi^3$-type interaction
Lagrangian density,
\be
{\cal L}=-g\phi\phi\Phi~,
\label{Lag_B}
\ee
by which we can get an estimation of $\Gamma_\phi$ for boson
constituent particle, $\phi$. 
\begin{figure}
\begin{center}
\includegraphics[scale=0.5]{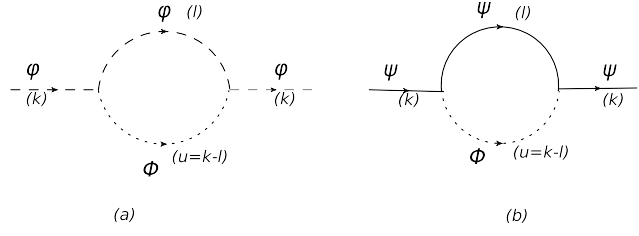}
\caption{Self-energy of $\phi$ boson for $\phi\Phi$ loop (a) 
and self-energy of $\psi$ fermion for $\psi\Phi$ loop (b).}
\label{gamma_B_F}
\end{center}
\end{figure}
The $\Phi$ is another higher mass boson like $\phi$.
The self-energy of $\phi$
for $\phi\Phi$ loop (as shown in Fig.~\ref{gamma_B_F}(a)) can be derived
in RTF with the help of the above
Lagrangian. 
Owing to the cutkosky
rule at finite $T$, the $\Gamma_\phi$ can be extracted from the imaginary part 
of this one-loop self-energy of $\phi$ at its pole, which is
already expressed in Eq.~(\ref{Gam_B}).

For estimating $\Gamma_\phi$, we have to repeat
a similar calculations as previously done in one-loop calculation 
for $\eta$. 
Therefore
Let us start
from 11 component of self-energy of $\phi$ for $\phi\Phi$ loop,
\be
\Pi^{11}_{(\phi)}(k)=-i\int \frac{d^4l}{(2\pi)^4}L(k,l)D_{11}(l,m_\phi)D_{11}(u=k-l,m_\Phi)~,
\label{Pi_11k}
\ee
where $D_{11}(l,m_\phi)$ and $D_{11}(u=k-l,m_\Phi)$ are the scalar
part of $\phi$ and $\Phi$ propagators 
at finite temperature with similar form like Eq.~(\ref{de11}).

Similar to Eq.~(\ref{Pi_LU}), we will get the imaginary part of retarded self-energy
for the $\phi\Phi$ loop,
\bea
{\rm Im}{\Pi}^{R}_{(\phi)}(k)&=&\pi\int\frac{d^3l}{(2\pi)^3}\frac{1}{4\om_l\om_u}
[L(l_0=\om_l,{\vec l},k)[\{1+n_l(\om_l)+n_u(k_0-\om_l)\}\delta(k_0 -\om_l-\om_u)
\nn\\
&&~~~~~~~~~~~+\{-n_l(\om_l)+n_u(-k_0+\om_l)\}\delta(k_0-\om_l+\om_u)]
\nn\\
&&+L(l_0=-\om_l,{\vec l},k)[\{n_l(\om_l)
-n_u(k_0+\om_l)\}\delta(k_0 +\om_l-\om_u)
\nn\\
&&~~~~~~~~~~~+\{-1-n_l(\om_l)-n_u(-k_0-\om_l)\}\delta(k_0 +\om_l+\om_u)]~,
\label{self_LU}
\eea
where $\om_u=\sqrt{({\vk}-{\vec l})^2+m_\Phi^2}$
and $n'$s are Bose-Einstein distribution functions. The region of 
the different branch cuts in $k_0$-axis are
($-\infty$ to $-\sqrt{\vk^2+(m_\Phi+m_\phi)^2}$) for unitary cut in negative $k_0$-axis,
($-\sqrt{\vk^2+(m_\Phi-m_\phi)^2}$ to $\sqrt{\vk^2+(m_\Phi-m_\phi)^2}$) for Landau cut and 
($\sqrt{\vk^2+(m_\Phi+m_\phi)^2}$ to $\infty$) for unitary cut in positive $k_0$-axis.
These are the different off-shell regions of $\phi$ boson, 
where imaginary part of its self-energy acquires a non-zero
values because of the different $\delta$ functions in Eq.~(\ref{self_LU}).
One can notice that the $\phi$ pole will remain within the Landau
cuts only when $m_\Phi>2m_\phi$. Hence we take an ad hoc
assumption $m_\Phi=3m_\phi$ to satisfy the threshold condition.

According to Eq.~(\ref{Gam_B}), the $\Gamma_\phi$ will be coming from 
Landau cut contribution associated with
the third term of Eq.~(\ref{self_LU}), which can be simplified as
\be
\Gamma_\phi=\frac{1}{16\pi\vk m_\phi}\int^{\tom^-}_{\tom^+}d\tom 
L(l_0=-\tom,{\vec l}=\sqrt{\tom^2-m_\phi^2},k_0=\om_k,\vk)
\{n_l(\tom)-n_u(\om_k+\tom)\}~,
\label{gm_BB}
\ee
where $\tom^{\pm}=\frac{R^2}{2m_\phi^2}(-\om_k\pm\vk W)$ with
$\om_k=\sqrt{\vk^2+m_\phi^2}$, 
$W=\sqrt{1-\frac{4m_\phi^4}{R^4}}$ and $R^2=2m_\phi^2-m_\Phi^2$.

Now to estimate $\Gamma_\psi$ of $\psi$ fermion, let us take a 
similar kind of interaction Lagrangian density,
\be
{\cal L}=-G{\ov \psi}\psi\Phi~,
\label{Lag_F}
\ee
where the mass of the $\Phi$ boson is again assumed
to be three times larger than the mass of the $\psi$ fermion
i.e. $m_\Phi=3m_\psi$. The self-energy of $\psi$, $\Sigma_{(\psi)}(k)$ for
$\psi\Phi$ loop is shown in Fig.~\ref{gamma_B_F}(b). 
%The imaginary part of its retarded component is given below
%\bea
%{\rm Im}{\Pi}^{R}_{(\psi)}(k)&=&\pi\int\frac{d^3l}{(2\pi)^3}\frac{1}{4\om_l\om_u}
%[L(l_0=\om_l,{\vec l},k)[\{1-n^+_l(\om_l)+n_u(k_0-\om_l)\}\delta(k_0 -\om_l-\om_u)
%\nn\\
%&&~~~~~~~~~~~+\{n^+_l(\om_l)+n_u(-k_0+\om_l)\}\delta(k_0-\om_l+\om_u)]
%\nn\\
%&&+L(l_0=-\om_l,{\vec l},k)[\{-n^-_l(\om_l)
%-n_u(k_0+\om_l)\}\delta(k_0 +\om_l-\om_u)
%\nn\\
%&&~~~~~~~~~~~+\{-1+n^-_l(\om_l)-n_u(-k_0-\om_l)\}\delta(k_0 +\om_l+\om_u)]~.
%\label{self_LU_F}
%\eea
%Again using Eq.~(\ref{Gam_F}), the third term among four
%$\delta$-functions is only contributed to obtain
%\be
%\Gamma_\psi=\frac{1}{16\pi\vk}\int^{\tom^-}_{\tom^+}d\tom 
%L(l_0=-\tom,{\vec l}=\sqrt{\tom^2-m_\psi^2},k_0=\om_k,\vk)
%\{-n^-_l(\tom)-n_u(\om_k+\tom)\}
%\label{gm_BF}
%\ee
According to Eq.~(\ref{Gam_F}), we get
\be
\Gamma_\psi=\frac{1}{16\pi\vk}\int^{\tom^-}_{\tom^+}d\tom 
L(l_0=-\tom,{\vec l}=\sqrt{\tom^2-m_\psi^2},k_0=\om_k,\vk)
\{n^-_l(\tom)+n_u(\om_k+\tom)\}~,
\label{gm_BF}
\ee
which is very similar to Eq.~(\ref{gm_BB}). 
The modified parameters are $\tom^{\pm}=\frac{R^2}{2m_\psi^2}(-\om_k\pm\vk W)$ with 
$\om_k=\sqrt{\vk^2+m_\psi^2}$, 
$W=\sqrt{1-\frac{4m_\psi^4}{R^4}}$ and $R^2=2m_\psi^2-m_\Phi^2$.
Here $n^-_l$ is Fermi-Dirac distribution function of anti-fermion
and $n_u$ is Bose-Einstein distribution function of $\Phi$ boson.

%Again for $\phi$ self-energy, there will be another possible
%$\psi\psi$ loop can be evaluated from the same Lagrangian of
%Eq.~(\ref{Lag_F}). This is shown in Fig.~\ref{gamma_B_F}(c).
%Here both distribution functions will be Fermi-Dirac distribution.
%Hence the thermal width of $\phi$ boson for $\psi\psi$ loop is
%\be
%\Gamma_\phi=\frac{-1}{16\pi\vk m_\phi}\int^{\tom^-}_{\tom^+}d\tom 
%L(l_0=-\tom,{\vec l}=\sqrt{\tom^2-m_\psi^2},k_0=\om_k,\vk)
%\{n^+_l(\tom)+n^+_u(\om_k+\tom)\}~.
%\label{gm_FF}
%\ee
%where $\tom^{\pm}=\frac{R^2}{2m_\phi^2}(-\om_k\pm\vk W)$ with 
%$W=\sqrt{1-\frac{4m_\psi^4}{R^4}}$ and $R^2=m_\phi^2$.
The vertex part $L(k,l)$ for these loops are respectively
given below
\be
L(k,l)=-g^2 ~~~({\rm for}~\phi\Phi~{\rm loop})~,
\ee
\be
L(k,l)=-G^2m_\psi~~~({\rm for}~\psi\Phi~{\rm loop})~.
\ee
%\bea
%L(k,l)&=&G^2{\rm Tr}[(\ls+m_\psi)(\ks+\ls+m_\psi)]~~~({\rm for}~\psi\psi~{\rm loop})
%\nn\\
%&=&4G^2(m_\psi^2+l^2+k\cdot l)~.
%\eea

\section{Higher order loop diagrams}
\begin{figure}
\begin{center}
\includegraphics[scale=0.5]{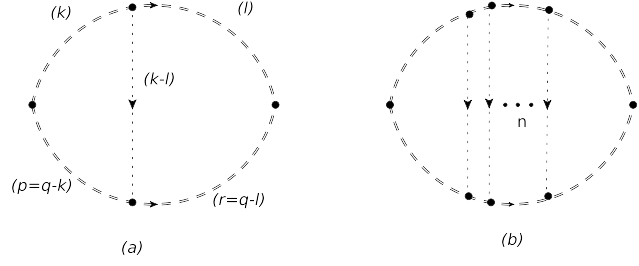}
\caption{two loops (a) and (n+1) loops (b)
ladder type diagrams for $\phi\phi\Phi$ interaction.}
\label{Ladder_kind}
\end{center}
\end{figure}

As $\Gamma_\phi\propto g^2$ (or $\Gamma_\psi\propto G^2$), hence
our one-loop contribution of two point viscous stress
tensor will be ${\cal O}(1/g^2)$ (since 
$\eta\propto{\cal O}(1/\Gamma_\phi)$).
Now there are some higher order loop diagrams which
are also ${\cal O}(1/g^2)$. 
First possible digram is shown in the Fig.~\ref{Ladder_kind}(a)
which contain two loops and two $\phi\phi\Phi$ vertex 
$v_{\phi\phi\Phi}\propto g$. Therefore its order will be 
${\cal O}(\frac{1}{\Gamma^2}v_{\phi\phi\Phi}^2)={\cal O}(1/g^2)$
(as $1/\Gamma$ will be coming from each loop).
Similarly we can get same order from the diagram, shown in the
Fig.~\ref{Ladder_kind}(b), which contain $(n+1)$ number of loops and $2n$ 
number of $v_{\phi\phi\Phi}$. Even for $n=\infty$ we get same order of
contribution.
\begin{figure}
\begin{center}
\includegraphics[scale=0.5]{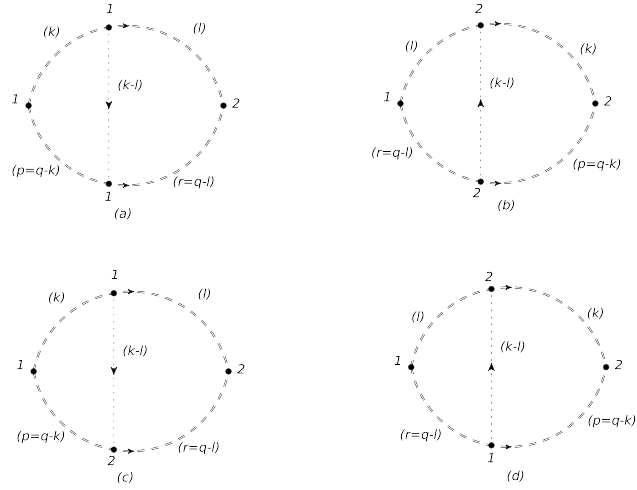}
\caption{The four possible combinations for two loop
diagrams in RTF. The
$\Pi^{AB}_{12}$ is the combinations of diagrams (a)
and (b) whereas diagrams (c) and (d) are combined in
$\Pi^{CD}_{12}$.}
\label{2loop_comp}
\end{center}
\end{figure}
\begin{figure}
\begin{center}
\includegraphics[scale=0.5]{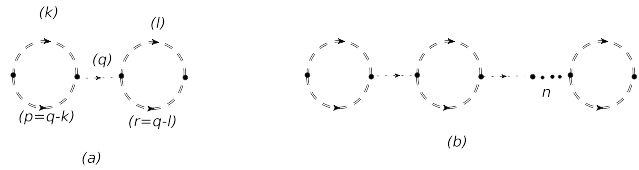}
\caption{two loops (a) and (n+1) loops (b)
rung type diagrams for $\phi\phi\Phi$ interaction.}
\label{Rung_kind}
\end{center}
\end{figure}
Let us first derive the two-loop digram (Fig.~\ref{Ladder_kind}(a)) 
to get a numerical estimation, which can be compared 
with the contributions of one-loop diagram. 
Instead of 11 component we will deduce the 12 component for
some certain advantage to simplify our calculation. As any one
of the four component in RTF can be
expressed in terms of the diagonal element ${\ov \Pi}$ due
to their relation
\be
\left(\begin{array}{cc} \Pi_{11} & \Pi_{12}
\\ \Pi_{21} & \Pi_{22} \end{array}\right)
=\left(\begin{array}{cc} \sqrt{1+n} & -\sqrt{n} 
\\ -\sqrt{n} & \sqrt{1+n} \end{array}\right)
\left(\begin{array}{cc} {\ov \Pi} & 0 
\\ 0 & -{\ov \Pi}^* \end{array}\right)
\left(\begin{array}{cc} \sqrt{1+n} & -\sqrt{n} 
\\ -\sqrt{n} & \sqrt{1+n} \end{array}\right)~.
\ee
Therefore we can express the shear viscosity in terms
of 12 component of two-loop skeleton digram as
\be
\eta^{(2)}=\frac{1}{10}\lim_{q_0,\vq \rightarrow 0}
\frac{\ep(q_0){\rm Im}{\ov\Pi}^{(2)}(q_0,\vq)}{q_0}
=\frac{1}{10}\lim_{q_0,\vq \rightarrow 0}
\frac{{\rm sinh}(\beta q_0/2)}{q_0}{\rm Im}\Pi^{(2)}_{12}(q_0,\vq)~.
\ee
Applying L'Hospital's rule, we have
\bea
\lim_{q_0 \rightarrow 0}\frac{{\rm sinh}(\beta q_0/2)}{q_0}&=&\lim_{q_0 \rightarrow 0}
\frac{\frac{d}{dq_0}\{{\rm sinh}(\beta q_0/2)\}}{\frac{d}{dq_0}\{q_0\}}
\nn\\
&=&\frac{\beta}{2}
\eea
and so
\be
\eta^{(2)}=\frac{\beta}{20}\lim_{q_0,\vq \rightarrow 0}{\rm Im}\Pi^{(2)}_{12}(q_0,\vq)~.
\label{eta2_Pi12}
\ee
Denoting the intermediate vertices $v_{\phi\phi\Phi}$ in four possible way as shown
in Fig.~(\ref{2loop_comp}), we get the 12 component as a summation of
four possible diagrams.
By attaching different momentum labels on the internal lines,
the four terms can be merged into two as given below~\cite{Baier}  
\be
i\Pi^{(2)}_{12}(q)=i\Pi^{AB}_{12}(q)+i\Pi^{CD}_{12}(q)~,
\ee
where 
\be
i\Pi^{AB}_{12}(q)=-i\int \frac{d^4kd^4l}{(2\pi)^8}N^{(2)}(q,k,l)
D^{\phi}_{12}(l) D^{\phi}_{21}(r=q-l) D^{AB}~,
\label{Pi_AB}
\ee
\be
i\Pi^{CD}_{12}(q)=i\int \frac{d^4kd^4l}{(2\pi)^8}N^{(2)}(q,k,l)
D^{\phi}_{12}(l) D^{\phi}_{21}(p=q-k)D^{\Phi}_{12}(u=k-l) D^{CD}~,
\label{Pi_CD}
\ee
with
\bea
D^{AB}&=&\{D^{\phi}_{11}(k) D^{\phi}_{11}(p=q-k) D^{\Phi}_{11}(u=k-l)+
D^{\phi}_{22}(k) D^{\phi}_{22}(p=q-k) D^{\Phi}_{22}(u=k-l)\}
\nn\\
&=&2i{\rm Im}\{D^{\phi}_{11}(k) D^{\phi}_{11}(p=q-k) D^{\Phi}_{11}(u=k-l)\}~,
\label{I_AB}
\eea
\bea
D^{CD}&=&\{D^{\phi}_{11}(k) D^{\phi}_{22}(r=q-l)+
D^{\phi}_{22}(k) D^{\phi}_{11}(r=q-l)\}
\nn\\
&=&2{\rm Re}\{D^{\phi}_{11}(k) D^{\phi}_{22}(r=q-l)\}~.
\label{I_CD}
\eea
The above simplifications have been done by using the identity $D_{11}=-D^*_{22}$.
The $N^{(2)}$ in the limit of $q_0,\vq\rightarrow 0$ can be expressed as
\be
N^{(2)}=-\frac{4}{3}g^2\vk^2\vl^2~,
\label{N2_kv}
\ee
where the angle between $\vk$ and $\vl$ is fixed
to zero for simplifying the calculations because
our main purpose in this section is just to
estimate the order of magnitude for the two-loop 
contribution in shear viscosity.

Now using the 12 component of scalar propagator,
\be
D^l_{12}=D^l_{21}=2\pi iF_l\delta(l_0^2-\om_l^2),~
F_l=\sqrt{n_l(1+n_l)}
\ee
in Eq.~(\ref{Pi_AB}), we have
\be
\Pi^{AB}_{12}(q)=\frac{-4i}{3}\int \frac{d^4l}{(2\pi)^4}(\vl^2)
[2\pi F_l\delta(l_0^2-\om_l^2)][2\pi F_r\delta((q_0-l_0)^2-\om_r^2)] \{I^{AB}\}~,
\ee
where 
\be
I^{AB}=2{\rm Im}G,~ G=\int\frac{d^4k(\vk^2g^2)}{(2\pi)^4}
\{D^{\phi}_{11}(k) D^{\phi}_{11}(p=q-k) D^{\Phi}_{11}(u=k-l)\}~.
\ee
After performing the $l_0$ integration, the Landau cut contribution
\footnote{As our main focus on $q_0,\vq\rightarrow 0$ for shear viscosity
calculations, so we have again considered only Landau cuts and excluded
the unitary cut contributions.}
of $\Pi^{AB}_{12}(q)$ is given by
\bea
\Pi^{AB}_{12}(q)&=&\frac{-4}{3}i
\int\frac{d^3l}{(2\pi)^3}\frac{\pi \vl^2F_lF_r}{2\om_l\om_r}
[I^{AB}(l_0=\om_l)\delta(q_0-\om_l+\om_r)
+I^{AB}(l_0=-\om_l)\delta(q_0+\om_l-\om_r)]
\nn\\
&=&\frac{-4}{3}i
\int\frac{d^3l}{(2\pi)^3}\frac{\vl^2F_lF_r}{2\om_l\om_r}
\lim_{\Gamma_\phi \rightarrow 0}
\left[\frac{\Gamma_\phi}{(q_0-\om_l+\om_r)^2+\Gamma_\phi^2}\{I^{AB}(l_0=\om_l)\}
\right.\nn\\
&&\left.~~~~+\frac{\Gamma_\phi}{(q_0+\om_l-\om_r)^2+\Gamma_\phi^2}\{I^{AB}(l_0=\om_l)\}\right]~.
\label{PiV_Gama}
\eea
Again starting with free theory ($\Gamma_\phi\rightarrow 0$), we will
make our calculations convert to interacting theory just by keeping
$\Gamma_\phi$ as a non-zero value.
In the limit of $q_0,\vq \rightarrow 0$, $\om_r\rightarrow\om_l$ and
$n_r\rightarrow n_l$. 
%In the above equation we have ignored
%the parts which are associated with ${\rm Re}\frac{1}{(q_0\pm\om_l\mp\om_r)+i\Gamma}$
%because they are vanished in the limit of $q_0,\vq \rightarrow 0$ for finite $\Gamma$.
So
\be
\lim_{q_0,\vq \rightarrow 0}\Pi^{AB}_{12}(q)=\frac{-4i}{3}\int\frac{d^3l}{(2\pi)^3}
\frac{\vl^2F_l^2}{2\om_l^2\Gamma_\phi}
\lim_{q_0,\vq \rightarrow 0}\{I^{AB}(l_0=\om_l)+I^{AB}(l_0=-\om_l)\}~.
\label{PiAB_I}
\ee
Now we concentrate on $I^{AB}=2{\rm Im}G$, where $G$ can be evaluated as
\bea
G&=&-\int\frac{d^3k}{(2\pi)^3}\frac{\vk^2g^2}{4\om_k\om_p}\{n_k(1+n_p)+n_p(1+n_k)\}
\left[\frac{\Gamma_\phi}{(q_0-\om_k+\om_p)^2+\Gamma_\phi^2}D^{\Phi}_{11}(u_0=\om_k-l_0)
\right.\nn\\
&&\left.~~~~+\frac{\Gamma_\phi}{(q_0+\om_k-\om_p)^2+\Gamma_\phi^2}
D^{\Phi}_{11}(u_0=-\om_k-l_0)\right]~.
\eea
Here also we have taken the Landau cut contribution for finite $\Gamma_\phi$ after
performing the $k_0$ integration. One can find that the calculation of $G$ is
very similar to the calculation of 11 component of self-energy for finite $\Gamma_\phi$
if we approximately assume $\vk^2D^{\Phi}_{11}$ as a vertex part.
In the limiting case,
\be
\lim_{q_0,\vq \rightarrow 0}G=-\int\frac{d^3k}{(2\pi)^3}
\frac{\vk^2g^2n_k(1+n_k)}{2\om_k^2\Gamma_\phi}[
D^{\Phi}_{11}(u_0=\om_k-l_0)+D^{\Phi}_{11}(u_0=-\om_k-l_0)]~.
\label{G_q0}
\ee
As 
\be
{\rm Im}D^{\Phi}_{11}(u)=\frac{\pi n_u(\om_u)}{\om_u}\{\delta(u_0-\om_u)+\delta(u_0+\om_u)\}~,
\ee
therefore
\bea
\lim_{q_0,\vq \rightarrow 0}I^{AB}(l_0)&=&-2\int\frac{d^3k}{(2\pi)^3}
\frac{\vk^2g^2n_k(1+n_k)}{2\om_k^2\Gamma_\phi}\frac{\pi n_u}{\om_u}
\{\delta(l_0-\om_k-\om_u)
\nn\\
&&+\delta(l_0-\om_k+\om_u)
+\delta(l_0+\om_k-\om_u)+\delta(l_0+\om_k+\om_u)\}~.
\label{PiV_allcut}
\eea
Here four delta functions provide four different regions in $l_0$-axis
to $I^{AB}(l_0)$, where the function will remain non-zero.
These regions are exactly similar to the non-zero regions of imaginary part
of any one-loop self-energy function (see Eq.~(\ref{self_LU})). 
For $l_0=\pm\om_l$, third and second delta functions respectively contribute
with same magnitude. Hence
\bea
\lim_{q_0,\vq \rightarrow 0}I^{AB}(l_0=\om_l)&=&\lim_{q_0,\vq \rightarrow 0}I^{AB}(l_0=-\om_l)
\nn\\
&=&\frac{-g^2}{4\pi\vl}\int^{\tom^-}_{\tom^+}d\tom 
\frac{(\tom^2-m_\pi^2)n_k(\tom)\{1+n_k(\tom)\}n_u(\om_l+\tom)}{\tom\Gamma_\phi}~,
\label{IV}
\eea
where $\tom^{\pm}=\frac{R^2}{2m_\phi^2}(-\om_l\pm\vl W)$ with 
$W=\sqrt{1-\frac{4m_\phi^4}{R^4}}$ and $R^2=2m_\phi^2-m_\Phi^2$. 
Using (\ref{IV}) in (\ref{PiAB_I}) and taking the imaginary part,
we have
\be
\lim_{q_0,\vq \rightarrow 0}{\rm Im}\Pi^{AB}_{12}(q)=\frac{4}{3}\int\frac{d^3l}{(2\pi)^3}
\frac{\vl^4n_l(1+n_l)}{2\om_l^2\Gamma_\phi}2F^{AB}(\vl,T)
\label{PiAB_F}
\ee
where
\bea
F^{AB}(\vl,T)&=&\frac{-1}{\vl^2}\lim_{q_0,\vq \rightarrow 0}I^{AB}(l_0=\om_l,\vl,T)
\nn\\
&=&\frac{g^2}{4\pi\vl^3}\int^{\tom^-}_{\tom^+}d\tom 
\frac{(\tom^2-m_\pi^2)n_k(\tom)\{1+n_k(\tom)\}n_u(\om_l+\tom)}{\tom\Gamma_\phi}~.
\label{F_AB}
\eea
Next to calculate $\Pi^{CD}$, we will first find
\bea
D^{CD}&=&2{\rm Re}\{D^{\pi}_{11}(k) D^{\pi}_{22}(r=q-l)\}
\nn\\
&=&2\left[\frac{-1}{k_0^2-\om_k^2+i\ep}\frac{1}{(q_0-l_0)^2-\om_r^2+i\ep}
\right.\nn\\
&&\left.+\{2i\pi n_k\delta(k_0^2-\om_k^2)\}\{2i\pi n_r\delta((q_0-l_0)^2-\om_r^2)\}\right]~,
\eea
in which 2nd part can be approximately ignored due to containing exponentially suppressing
term $n_kn_r$. Using its first part in Eq.~(\ref{Pi_CD}), we have
\be
\Pi^{CD}_{12}(q)=2\int \frac{d^4k}{(2\pi)^4}\frac{(-4\vk^2)}{3}
\left[\frac{-1}{k_0^2-\om_k^2+i\ep}
2i\pi F_p\delta((q_0-k_0)^2-\om_p^2)\right]I^{CD}~,
\label{Pi_CD_Ik}
\ee
where
\be
I^{CD}=\int \frac{d^4l}{(2\pi)^4}(g^2\vl^2)
\left[\frac{D^\sigma_{12}(k-l)}{(q_0-l_0)^2-\om_r^2+i\ep}2i\pi F_l\delta(l_0^2-\om_l^2)\right]~.
\label{ICDk}
\ee
After doing $k_0$ integration of Eq.~(\ref{Pi_CD_Ik}) and taking the Landau
cut contribution with finite $\Gamma_\phi$ (in place of $\ep$ in (\ref{Pi_CD_Ik}))
we will get
\be
\lim_{q_0,\vq \rightarrow 0}\Pi^{CD}_{12}(q)=\frac{4}{3}
\int\frac{d^3k}{(2\pi)^3}\frac{\vk^2F_k}{2\om_k^2\Gamma_\phi}
\lim_{q_0,\vq \rightarrow 0}\{I^{CD}(k_0=\om_k)+I^{CD}(k_0=-\om_k)\}~,
\label{PiCD_I}
\ee
which is very similar to Eq.~(\ref{PiAB_I}). Going through the similar process
for Eq.~(\ref{ICDk}), we have
\be
\lim_{q_0,\vq \rightarrow 0}I^{CD}(k_0)=\int\frac{d^3l}{(2\pi)^3}
\frac{g^2\vl^2F_l}{2\om_l^2\Gamma_\phi}\{D^\Phi_{12}(u_0=k_0-\om_l)
+D^\Phi_{12}(u_0=k_0+\om_l)\}~,
\label{ICD_D12}
\ee
which can be compared with (\ref{G_q0}). 
Now the imaginary part of Eq.~(\ref{PiCD_I}) is
\be
\lim_{q_0,\vq \rightarrow 0}{\rm Im}\Pi^{CD}_{12}(q)=\frac{4}{3}
\int\frac{d^3k}{(2\pi)^3}\frac{\vk^2F_k}{2\om_k^2\Gamma_\phi}
\lim_{q_0,\vq \rightarrow 0}\{{\rm Im}I^{CD}(k_0=\om_k)
+{\rm Im}I^{CD}(k_0=-\om_k)\}~,
\label{PiCD_I_im}
\ee
where
\bea
{\rm Im}I^{CD}(k_0=\om_k,\vk)&=&{\rm Im}I^{CD}(k_0=-\om_k,\vk)
\nn\\
&=&\frac{g^2}{16\pi\vk}\int^{\tom^-}_{\tom^+}d\tom 
\frac{(\tom^2-m_\phi^2)F_l(\tom)F_u(\om_k+\tom)}{\tom\Gamma_\phi}~.
\label{ICDV}
\eea
%where $\tom^{\pm}=\frac{R^2}{2m_\phi^2}(-\om_k\pm\vk W)$ with 
%$W=\sqrt{1-\frac{4m_\phi^4}{R^4}}$ and $R^2=2m_\phi^2-m_\Phi^2$.
One can notice that the Eq.~(\ref{ICDV}) is very analogous
to (\ref{IV}).
Using (\ref{ICDV}) in (\ref{PiCD_I_im}), we have
\be
\lim_{q_0,\vq \rightarrow 0}{\rm Im}\Pi^{CD}_{12}(q)=\frac{4}{3}\int\frac{d^3k}{(2\pi)^3}
\frac{\vk^4n_k(1+n_k)}{2\om_k^2\Gamma_\phi}2F^{CD}(\vk,T)~,
\label{PiCD_F}
\ee
where
\bea
F^{CD}(\vk,T)&=&\frac{1}{F_k\vk^2}\lim_{q_0,\vq \rightarrow 0}I^{CD}(k_0=\om_k,\vk,T)
\nn\\
&=&\frac{g^2}{16\pi F_k\vk^3}\int^{\tom^-}_{\tom^+}d\tom 
\frac{(\tom^2-m_\pi^2)F_l(\tom)F_u(\om_k+\tom)}{\tom\Gamma_\phi}~.
\label{F_CD}
\eea
Using (\ref{PiAB_F}) and (\ref{PiCD_F}) in (\ref{eta2_Pi12}) we get the 
final expression of shear viscosity for two loop diagram as
\be
\eta^{(2)}=\frac{\beta}{30\pi^2}\int\frac{d\vk F(\vk,T)\vk^6}
{\Gamma_\phi\om_k^2}n_k(1+n_k)~,
\label{eta2_final}
\ee
where
\be
F(\vk,T)=F^{AB}(\vk,T)+F^{CD}(\vk,T)
\label{F_ABCD}
\ee
is the main factor which makes the $\eta^{(2)}$ be different from the $\eta^{(1)}$ 
(let us denote $\eta$ of Eq.~(\ref{eta_BB}) by $\eta^{(1)}$ to mark as 
one-loop contribution)
since Eq.~(\ref{eta2_final}) become identical with Eq.~(\ref{eta_BB}) for
$F(\vk,T)=1$.
%From the Fig.~(\ref{eta2}) we see that $\eta^{(2)}$ is almost
%$95\%$ smaller than $\eta^{(1)}$ in the temperature range ($T=0.1-0.175$
%GeV), in which the (net) baryon free hadronic matter is supposed to be evolved
%in the laboratories of heavy ion collision (HIC). On the basis of this numerical 
%comparision, we may safely ignore the two loops as well as the higher loops 
%contributions. 

Another pattern of two loop diagram and its nth extension are shown in Fig.~\ref{Rung_kind}(a)
and (b) respectively. Though these are seems to be contributed with ${\cal O}(1/\Gamma_\phi)$ 
but actually they can be analytically obtained as zero.
The two loop diagram in the static limit can be expressed as
\bea
\lim_{q_0,\vq \rightarrow 0}\Pi^{(2)}_{11}&=&-\lim_{q_0,\vq \rightarrow 0}
\left[\int \frac{d^4k}{(2\pi)^4}N(q,k) D^{\phi}_{11}(k) D^{\phi}_{11}(p=q-k)\right]
D^\Phi_{11}(q)
\nn\\
&&\left[\int \frac{d^4l}{(2\pi)^4}N(q,l) D^{\phi}_{11}(l) D^{\phi}_{11}(p=q-l)\right]
\nn\\
&=&-\left[i\int \frac{d^3k}{(2\pi)^3}\frac{-N(\vk)n_k(1+n_k)}{\Gamma_\phi\om_k^2}\right]
\left[\lim_{q_0,\vq \rightarrow 0}\{{\rm Re}D^\Phi_{11}+i{\rm Im}D^\Phi_{11}\}\right]
\nn\\
&&\left[i\int \frac{d^3l}{(2\pi)^3}\frac{-N(\vl)n_l(1+n_l)}{\Gamma_\phi\om_l^2}\right]~.
\eea
Its imaginary part is directly proportional to 
\be
\lim_{q_0,\vq \rightarrow 0}{\rm Im}D^\Phi_{11}=\lim_{q_0,\vq \rightarrow 0}
2\pi n^\Phi_q\delta(q^2-m_\Phi^2)~,
\ee
which is exactly equal to zero.
Therefore $\eta^{(2)}$, which is related with imaginary part of $\Pi^{(2)}_{11}$,
is also become zero.
\section{Numerical discussion}
\begin{figure}
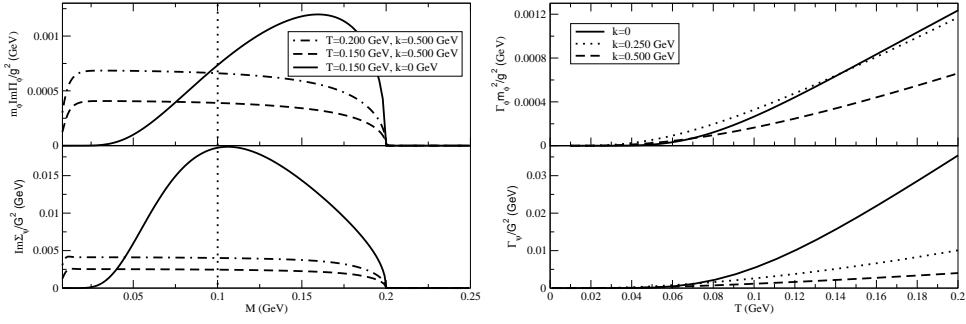

\begin{center}
\includegraphics[scale=0.25]{im_M.eps}
\includegraphics[scale=0.25]{G_T.eps}
\caption{Left : Imaginary part of self-energy of $\phi$ boson
(upper panel) and $\psi$ fermion (lower panel) for different values
of three-momentum $\vk$ and temperature $T$. The dotted line is indicating
the pole positions of $\phi$ boson or $\psi$ fermion 
($m_\phi=m_\psi=0.100$ GeV).
Right : the $T$ dependence of thermal widths ($\Gamma_\phi$ and $\Gamma_\psi$)
for $\phi$ (upper panel) and $\psi$ (lower panel).}
\label{im_M_GT}
\end{center}
\end{figure}
\begin{figure}
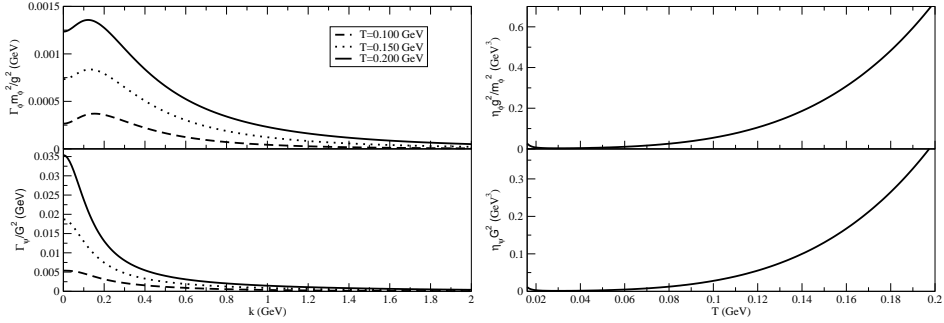

\begin{center}
\includegraphics[scale=0.25]{G_k.eps}
\includegraphics[scale=0.25]{Eta_T.eps}
\caption{Left : $\Gamma_\phi$ (upper panel) and $\Gamma_\psi$ (lower panel) 
vs ${\vec k}$.
Right : $T$ dependence of shear viscosity coefficients $\eta_\phi$ 
and $\eta_\psi$ for $\phi$ boson (upper panel) and $\psi$ fermion 
(lower panel) respectively.}
\label{Gk_eta_T}
\end{center}
\end{figure}
\begin{figure}
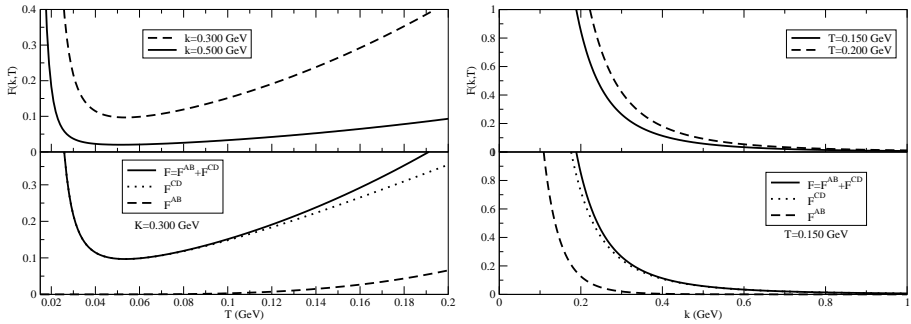

\begin{center}
\includegraphics[scale=0.25]{F_T.eps}
\includegraphics[scale=0.25]{F_Kv.eps}
\caption{Upper panel : $T$ (left) and $\vk$ (right) dependency
of $F(\vk,T)$.
Lower panel : $T$ (left) and $\vk$ (right) dependency
of individual contributions of $F^{AB}$ (dashed line), 
$F^{CD}$ (dotted line) and their total $F$ (solid line).}
\label{F_T_k}
\end{center}
\end{figure}
\begin{figure}
\begin{center}
\includegraphics[scale=0.5]{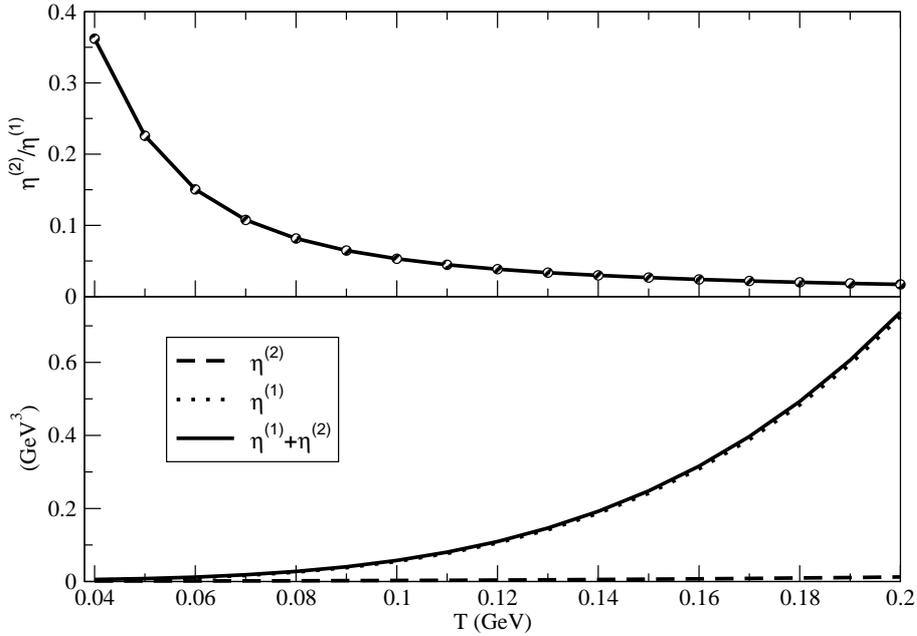}
\caption{Upper panel : The
variation of ratio of two-loop ($\eta^{(2)}$) to one-loop
($\eta^{(1)}$) contributions in shear viscosity with
temperature $T$.
Lower panel : $T$ dependency of the individual contribution of $\eta^{(2)}$
(dashed line), $\eta^{(1)}$ (dotted line) and their total (solid line).}
\label{eta2}
\end{center}
\end{figure}
Let us first concentrate on the numerical results of thermal
width and then on the results of shear viscosity coefficient.
In Sec.(3), we have noticed that thermal width of $\phi$ boson and
$\psi$ fermion 
are coming from the Landau cut contributions of their corresponding
one-loop self-energies which are shown in the left panel of
Fig.~(\ref{im_M_GT}). Here the imaginary part of the $\phi$ (upper
panel) and $\psi$ (lower panel) self-energies are exhibiting their
non-zero contributions within the Landau cut regions
(from 0 to ($m_\Phi-m_{\phi,\psi}$) i.e. from 0 to $0.2$ GeV) 
in their invariant mass axes, $M$.
We have taken $m_\phi=m_\psi=0.1$ GeV for making a comparative analysis.
Thermal widths $\Gamma_\phi$ and $\Gamma_\psi$ for $\phi$ 
and $\psi$, respectively, are basically the contributions at their corresponding pole
positions ($M=m_\phi$ and $m_\psi$), which are marked by dotted line
in the left panel of Fig.~(\ref{im_M_GT}). The $\Gamma_\phi$ and $\Gamma_\psi$ 
increase with increasing temperature ($T$) and decreasing
three-momentum (${\vec k}$). These are shown in the right
panel of Fig.~(\ref{im_M_GT}) and left panel of Fig.~(\ref{Gk_eta_T}) respectively.
We have presented the normalized values, $m_\phi^2\Gamma_\phi/g^2$
and $\Gamma_\psi/G^2$ to make our descriptions more general for
any arbitrary value of coupling constant, $g$ or $G$. 
Now, using these thermal widths $\Gamma_\phi(\vk,T)$ and $\Gamma_\psi
(\vk,T)$ in Eqs.~(\ref{eta_BB}) and (\ref{eta_FF}), we get the 
shear viscosity coefficients $\eta_\phi$ and $\eta_\psi$ respectively.
Their corresponding temperature dependence are shown in upper
and lower panels of Fig.~(\ref{Gk_eta_T}) on the right side.
Here we have again normalized our results for unit coupling
constants and so $g^2\eta_\phi/m_\phi^2$ vs $T$ (upper panel) 
and $G^2\eta_\psi$ vs $T$ (lower panel) have been represented in this graph.
In Sec.(4), we have pointed 
out the possibility of infinite number of 
ladder-type diagrams (shown in Fig.~\ref{Ladder_kind}) which are
supposed to be of same order of magnitude (${\cal O}(1/g^2)$) like
the one-loop (shown in Fig.~\ref{diagram}) contributions. 
This is very well known complexity for the calculation of 
transport coefficient via Kubo approach.
A two loop calculation for boson self-energy is derived
to check its numerical strength with respect to the one-loop contribution.
In the expression for the two-loop, a quantity $F(\vk,T)$ in the Eq.~(\ref{eta2_final}) 
is recognized as the main factor to differentiate between the 
two-loop and the one-loop expression.
The contributions of $F^{AB}(\vk,T)$, $F^{CD}(\vk,T)$
and their total $F(\vk,T)$ from Eq.~(\ref{F_ABCD}) are
shown in the lower panel of the left and right side of Fig.~(\ref{F_T_k}).
The $F^{CD}(\vk,T)$ is numerically so stronger than $F^{AB}(\vk,T)$
that one can confidently assume $F(\vk,T)\simeq F^{CD}(\vk,T)$.
The $T$ and $\vk$ dependence of $F(\vk,T)$ are shown in the upper
panel of left and right side of Fig.~(\ref{F_T_k}) for two different
values of fixed $\vk$ and $T$ respectively. Using the $F(\vk,T)$ in
Eq.~(\ref{eta2_final}) we have obtained the normalized values of $\eta^{(2)}$
(i.e. $\eta^{(2)}g^2/m_\phi^2$) which are displayed
by dashed line in the lower panel of Fig.~(\ref{eta2}). 
The ratio $\eta^{(2)}/\eta^{(1)}$ vs $T$  is presented
in the upper panel of Fig.~(\ref{eta2}).
The figure shows that $\eta^{(2)}$ is substantially suppressed from
the one-loop contribution $\eta^{(1)}$. 
During extension from one-loop to two-loop calculation, 
the additional
thermal distribution functions are introduced in multiplicative
way. This may be the main reason of the suppression of $\eta^{(2)}$
from $\eta^{(1)}$. This two loop result is indicating that
as the number of loops will be increased, its numerical strength
will be successively suppressed.
So the one-loop contribution of the shear viscosity for this 
simple $\phi\phi\Phi$ interaction may be considered as
a leading contribution. 
This approximation is valid especially for the temperature
greater than the mass of constituent
particles of the medium (i.e. $T>m_\phi$).
In some sense this  estimation
is very general as it is independent of coupling constant $g$.
\section{Summary and conclusion}
Taking a special and simple $\phi\phi\Phi$ or $\psi\psi\Phi$ interaction
Lagrangian, a diagrammatic analysis of Kubo-type shear viscosity
has been presented in this manuscript. At first the simplest possible skeleton, one-loop
diagram for $\phi$ boson (or $\psi$ boson) is evaluated in real-time
thermal field theory, where the effective propagators with finite thermal
width are used in the $\phi\phi$ or $\psi\psi$ loop. The thermal
width of $\phi$ is extracted from the imaginary part of its thermal
self-energy for $\phi\Phi$ or $\psi\Phi$ loop. After the one-loop
analysis, the higher order loop diagrams with same power of
coupling constant as in the one-loop are inspected. Instead
of re-summing them, the next possible skeleton, two-loop
diagram is explicitly deduced in RTF.
Extending the calculations from one-loop to two-loop, the extra
thermal distribution function comes automatically into the 
picture, for which the numerical strength of two-loop is suppressed
from one-loop contribution. It is naturally expected that
as we increase the number of loops, the suppression will
successively grow. On this basis, the one-loop results
may be considered as a leading order results for this simple
$\phi\phi\Phi$ interaction. 
As a practical example, if someone is interested to calculate the shear
viscosity of hot pionic medium~\cite{GKS} by using the $\pi\pi\sigma$  
as well as $\pi\pi\rho$ interaction (effective) Lagrangian,
then one-loop estimation is sufficient for numerical purpose.
Again, we know that this Kubo-type one-loop 
expression of shear viscosity in terms of thermal width
coincides exactly with the expression from the relaxation time approximation of
kinetic theory. Hence, one can get similar results
for the $\phi\phi\Phi$ or $\psi\psi\Phi$ interaction following
relaxation time approximation also.

\section*{Acknowledgments}Author thanks to Prof. Sourav Sarkar and 
Prof. Angel Gomez Nicola for their useful help and suggestions. 
Author also thanks to Abhishek Mishra, Sandeep Gautam, 
Supriya Mondal for their useful suggestions
to improve the writings of this article.
This work is financed by Fundacao de Amparo a Pesquisa do Estado de Sao Paulo 
(FAPESP) under Contract No. 2012/16766-0.

\appendix

\section{Calculation of $\pi^{\mn}$ :}
Using the free Lagrangian densities
\be
{\cal L}=\frac{1}{2}\del_\mu\phi\del^\mu\phi -\frac{1}{2}m_\phi^2\phi^2
\ee
and
\be
{\cal L}={\ov \psi}(i\gamma^\mu\del_\mu -m_\psi)\psi
\ee
for $\phi$ boson and $\psi$ fermion,
their energy momentum tensors can respectively be obtained as
\bea
T_{\rho\sigma}&=&-g_{\rho\sigma}{\cal L}
+\frac{\del {\cal L}}{\del(\del^\rho\phi)}\del_\sigma\phi
+\frac{\del {\cal L}}{\del(\del^\sigma\phi)}\del_\rho\phi
\nn\\
&=&-g_{\rho\sigma}{\cal L}+\del_\sigma\phi\del_\rho\phi
\eea
and
\bea
T_{\rho\sigma}&=&-g_{\rho\sigma}{\cal L}
+\frac{\del {\cal L}}{\del(\del^\rho\psi)}\del_\sigma\psi
\nn\\
&=&-g_{\rho\sigma}{\cal L}+i{\ov \psi}\gamma_\rho\del_\sigma\psi~.
\eea
Now, the viscous stress tensor is defined as
\be
\pi_{\mn}=t^{\rho\sigma}_{\mn}T_{\rho\sigma}~,
\ee
where 
\be
t^{\rho\sigma}_{\mn}=\Delta^\rho_\mu\Delta^\sigma_\nu
-\frac{1}{3}\Delta_{\mn}\Delta^{\rho\sigma}~.
\ee
Using the relation $t^{\rho\sigma}_{\mn}g_{\rho\sigma}{\cal L}=0$, the 
$\pi^{\mn}$ for $\phi$ and $\psi$ can respectively be written as
\be
\pi_{\mn}=(\Delta^\rho_\mu\Delta^\sigma_\nu-\frac{1}{3}\Delta_{\mn}\Delta^{\rho\sigma})
\left\{\begin{array}{ll}
\del_\rho\phi\del_\sigma\phi
\\
i{\ov \psi}\gamma_\rho\del_\sigma\psi
\end{array}
\right\}~.
\label{pimn_Ap}
\ee
\section{Calculation of $N(q,k)$ :}
Let us write the 11-component of two point function of viscous
stress tensor in terms of field operators. For $\phi$ field
it is given by
\bea
\Pi_{11}(q)&=&i\int d^4x e^{iqx}\langle T\pi_{\alpha\beta}(x)\pi^{\alpha\beta}(0)\rangle_\beta
\nn\\
&=&t^{\rho\sigma}_{\alpha\beta}t^{\alpha\beta}_{\mn}i\int d^4x e^{iqx}\langle T\del_\rho\phi(x)\del_\sigma\phi(x)
\del^\mu\phi(0)\del^\nu\phi(0)\rangle_\beta~.
\label{pi_ab_Ap}
\eea
With the help of the Wick's contraction technique, we have
\bea
\Pi_{11}(q)&=&t^{\rho\sigma}_{\alpha\beta}t^{\alpha\beta}_{\mn}i\int d^4x e^{iqx}
[\langle T\del_\sigma\phi\underbrace{(x)\del_\rho\phi\overbrace{(x)
\del^\mu\phi}(0)\del^\nu\phi}(0)\rangle_\beta
\nn\\
&&+\langle T\del_\rho\phi\underbrace{(x)\del_\sigma\phi\overbrace{(x)
\del^\mu\phi}(0)\del^\nu\phi}(0)\rangle_\beta]
\nn\\
&=&t^{\rho\sigma}_{\alpha\beta}t^{\alpha\beta}_{\mn}
i\int \frac{d^4k}{(2\pi)^4}[N^{\mn}_{\rho\sigma}(q,k)+N^{\nu\mu}_{\rho\sigma}(q,k)] 
D_{11}(k)D_{11}(p=q-k)
\nn\\
&=&i\int \frac{d^4k}{(2\pi)^4}N(q,k) 
D_{11}(k)D_{11}(p=q-k)~,
\eea
where 
\bea
N(q,k)&=&t^{\rho\sigma}_{\alpha\beta}t^{\alpha\beta}_{\mn}
[N^{\mn}_{\rho\sigma}(q,k)+N^{\nu\mu}_{\rho\sigma}(q,k)]
\nn\\
&=&t^{\rho\sigma}_{\mn}
[N^{\mn}_{\rho\sigma}(q,k)+N^{\nu\mu}_{\rho\sigma}(q,k)]~.
\eea
From the one-loop self-energy graph of Fig.~(\ref{diagram}), the 
$N^{\mn}_{\rho\sigma}$ can be obtained as
\be
N^{\mn}_{\rho\sigma}=-k_\rho(q-k)_\sigma k^\mu(q-k)^\nu
\ee
and so in the comoving frame (i.e. $u=1,{\vec 0}$), we get
\be
N(q,k)=-[\vk^2(\vq-\vk)^2+\frac{\{\vk\cdot(\vq-\vk)\}^2}{3}]~.
\ee
Similarly for $\psi$ field,
\bea
\Pi_{11}(q)&=&t^{\rho\sigma}_{\alpha\beta}t^{\alpha\beta}_{\mn}i\int d^4x e^{iqx}
\langle T{\ov \psi}\underbrace{(x)\gamma_\rho\del_\sigma\psi\overbrace{(x)
{\ov \psi}}(0)\gamma^\mu\del^\nu\psi}(0)\rangle_\beta
\nn\\
&=&-i\int \frac{d^4k}{(2\pi)^4}N(q,k) 
D_{11}(k)D_{11}(p=q+k)~,
\eea
where 
\bea
N(q,k)&=&t^{\rho\sigma}_{\mn}N^{\mn}_{\rho\sigma}(q,k)
\nn\\
&=&t^{\rho\sigma}_{\mn}{\rm Tr}[\gamma^\mu(q+k)^\nu(\qs+\ks+m_p)
\gamma_\rho k_\sigma(\ks+m_k)]~.
\eea
In the comoving frame the $N(q,k)$ become
\be
N(q,k)=\frac{32}{3}\{k_0(q_0+k_0)\}\{\vk\cdot(\vq+\vk)\}
-4[\{\vk\cdot(\vq+\vk)\}^2+\frac{\vk^2(\vq+\vk)^2}{3}]~.
\ee
%\be 
%N(q,k)={\rm Tr}\{v_{1\mn}(\ks + m)v_2^{\mn}(\ks +\qs +m)\}
%\ee
%where
%\bea
%v_{1\mn}(k)&=&i(\Delta^\rho_\mu\Delta^\sigma_\nu-\frac{1}{3}\Delta_{\mn}\Delta^{\rho\sigma})
%(i\gamma_\rho)(-ik_\sigma)
%\nn\\
%&=&i[\{\gamma_\mu-\us u_\mu\}\{k_\nu-(k\cdot u)u_\nu\}
%\nn\\
%&&-\frac{1}{3}\{\ks - (k\cdot u)\us\}\Delta_{\mn}]
%\nn\\
%\eea
%and $v_2^{\mn}$ are same as $v_1^{\mn}$ but only the momentum $k$ is
%replaced by $(k+q)$. After going through the straight forward trace 
%algebra, we will get (in the comoving frame i.e. $u=1,{\vec 0}$)
%\be
%N(q,k)=\frac{16}{3}\vk\cdot\vp[2k^0p^0-2m^2-\vk\cdot\vp]
%\label{N_qvkv_FF}
%\ee
%Similarly for BB loop, we will get
%\be 
%N(q,k)=w_{1\mn}w_2^{\mn}
%\ee
%where
%\bea
%w_{1\mn}(k)&=&i(\Delta^\rho_\mu\Delta^\sigma_\nu-\frac{1}{3}\Delta_{\mn}\Delta^{\rho\sigma})
%(ik_\rho)(ip_\sigma)
%\nn\\
%&=&i[\{k_\mu-(u\cdot k) u_\mu\}\{p_\nu-(p\cdot u)u_\nu\}
%\nn\\
%&&-\frac{1}{3}\{(k\cdot p) - (k\cdot u)(p\cdot u)\}\Delta_{\mn}]
%\nn\\
%\eea
%and $w_2^{\mn}$ are exactly same as $w_1^{\mn}$.
%In the comoving frame ($u=1,{\vec 0}$), the results for BB loop become
%\be
%N(q,k)=[\vk^2\vp^2-\frac{1}{3}(\vk\cdot\vp)^2]
%\label{N_qvkv_BB}
%\ee
%\section*{References}

\end{document}